\documentclass[twocolumn,aps]{revtex4}
\usepackage{graphicx}
\usepackage{amssymb}
\usepackage{amsmath}
\usepackage{bm}
\usepackage{color}

\def\e{\varepsilon}
\def\ef{\varepsilon_{\rm F}}

\def\s{\sigma}
\def\t{\theta}

\def\ca{\chi_{\rm intra}}
\def\ce{\chi_{\rm inter}}
\def\bk{{\bm k}}
\def\mb{\mu_{\rm B}}
\def\kc{k_{\rm c}}
\def\ec{\varepsilon_{\rm c}}

\begin{document}

\title{Spin susceptibility of three-dimensional Dirac semimetals}
\author{Yuya Ominato and Kentaro Nomura}
\affiliation{Institute for Materials Research, Tohoku University, Sendai 980-8577, Japan}
\date{\today}

\begin{abstract}
We theoretically study the spin susceptibility of Dirac semimetals using the linear response theory. The spin susceptibility is decomposed into an intraband contribution and an interband contribution. We obtain analytical expressions for the intraband and interband contributions of massless Dirac fermions. The spin susceptibility is independent of the Fermi energy while it depends on the cutoff energy, which is introduced to regularize the integration. We find that the cutoff energy is appropriately determined by comparing the results for the Wilson-Dirac lattice model, which approximates the massless Dirac Hamiltonian around the Dirac point. We also calculate the spin susceptibility of massive Dirac fermions for the model of topological insulators. We discuss the effect of the band inversion and the strength of spin-orbit coupling.
\end{abstract}
\maketitle

\section{Introduction}
\label{intro}

Topological semimetals, such as Dirac semimetals \cite{young2012dirac,wang2012dirac}, Weyl semimetals  \cite{murakami2007phase,burkov2011weyl,wan2011topological}, and nodal line semimetals \cite{burkov2011topological,phillips2014tunable,kim2015dirac,yu2015topological}, possess exotic electronic band structure, which is significantly different from conventional metals and insulators. They exhibit fascinating physical properties originating from their topologically nontrivial band structure.
There are many theoretical proposals to realize topological semimetals, some of which were experimentally confirmed \cite{liu2014discovery,neupane2014observation,borisenko2014experimental,xu2015discovery,lu2015experimental,lv2015experimental}.
A Dirac semimetal has band touching points and the energy bands are doubly degenerate. By breaking either inversion symmetry or time-reversal symmetry, the degeneracy is lifted and a Dirac semimetal becomes a Weyl semimetal. The inversion broken Weyl semimetals are experimentally confirmed \cite{xu2015discovery,lu2015experimental,lv2015experimental} and there are several materials including type II Weyl semimetals \cite{deng2016experimental}.
On the other hand, there are few experimental indications for the Weyl semimetals with broken time-reversal symmetry, i.e. the magnetic Weyl semimetals \cite{nakatsuji2015large,nayak2016large,liu2017giant}, though there are many theoretical predictions \cite{burkov2011topological,wan2011topological,kurebayashi2014weyl,wang2016time,ito2017anomalous,yang2017topological,jin2017ferromagnetic,Xu2017topological,cho2011possible}.

One of the theoretical predictions to realize the magnetic Weyl semimetals is magnetically doped topological insulators \cite{yu2010quantized,kurebayashi2014weyl,cho2011possible,liu2013chiral}.
Ferromagnetic ordering in topological insulators is experimentally observed \cite{chen2010massive,wray2011topological,liu2012crossover,zhang2012interplay,zhang2013topology,chang2013experimental,chang2013thin}. In these systems, the ferromagnetic Weyl phase can emerge if the exchange coupling is sufficiently strong to overcome the energy gap.
The magnetic properties and the topological phase transition induced by magnetic doping are characterized by the spin susceptibility of band electrons.
Within the mean field theory, a condition to exhibit the ferromagnetic ordering is given by $J^2\chi_m\chi_s>1$ \cite{yu2010quantized}, where $J$ is the exchange coupling constant, $\chi_m$ is the spin susceptibility of local magnetic moments, and $\chi_s$ is the spin susceptibility of band electrons.
$\chi_m$ obeys the Curie law and is proportional to inverse of temperature ($\chi_m\propto 1/T$).
Therefore, the ferromagnetic ordering can be observed at sufficiently low temperature as long as $\chi_s$ is finite.
The investigation of $\chi_s$ in topological semimetals and insulators is an important issue to discuss the magnetic and topological phase transition in these systems.

In this paper, we study the spin susceptibility of three-dimensional Dirac semimetals within the linear response theory. 
The spin susceptibility is composed of the intraband contribution $\chi_{\rm intra}$ and the interband contribution $\chi_{\rm inter}$.
In the presence of strong spin-orbit coupling, $\chi_{\rm inter}$ gives large contribution.
The interband effect is important in the orbital diamagnetism of the Dirac fermions \cite{fukuyama1970interband,fuseya2009interband,fuseya2012spin,koshino2016magnetic}.
We obtain analytical expressions for the spin susceptibility of the massless Dirac fermions.
The spin susceptibility is independent of Fermi energy while it depends on the cutoff energy, which is introduced by hand to regularize the integration.
We calculate the spin susceptibility of the Wilson-Dirac lattice model, which reduces to the massless Dirac Hamiltonian around the $\Gamma$ point.
We find that the cutoff energy can be related to some parameters of the lattice model and 
that the Fermi energy dependence of the spin susceptibility exhibits quantitatively the same behavior in the two models.
We also calculate the spin susceptibility of massive Dirac fermions, which are models of band electrons in topological insulators.
The spin susceptibility is finite even in the energy gap because of strong spin-orbit coupling.

The paper is organized as follows. In Sec.\ \ref{sec_spin}, we formulate the spin susceptibility and briefly review qualitative behavior of the spin susceptibility in the presence of spin-orbit coupling.
In Sec.\ \ref{sec_continuum} and \ref{sec_lattice}, we introduce a continuum model and a lattice model which describe electronic states in a Dirac semimetal. The spin susceptibility of them is calculated. In Sec.\ \ref{sec_ti}, we calculate the spin susceptibility of massive Dirac fermions.
The conclusion is given in Sec.\ \ref{sec_conclusion}.

\section{spin susceptibility}
\label{sec_spin}

To calculate the spin susceptibility, we introduce the Zeeman coupling between the electrons and an external magnetic field. The Hamiltonian is given by
\begin{align}
H=H_0+H_{\rm Zeeman}
\end{align}
where $H_0$ is an unperturbed Hamiltonian and the Zeeman term is given by
\begin{align}
H_{\rm Zeeman}=\frac{g\mb}{2}\bm{\s}\cdot\bm{B},
\end{align}
where $g$ is the $g$ factor, $\mb$ is the Bohr magneton, and $\bm{\s}$ is the triplets of Pauli matrices acting on the real spin degree of freedom.

We apply an external magnetic field with infinitely slow spatial variation
\begin{align}
\bm{B}=\left(0,0,B\cos(\bm{q}\cdot\bm{r})\right).
\end{align}
The slow spatial variation of the field is controlled by the wave vector $\bm{q}$, which will tend to zero at the end of the calculation. Within the linear response, the induced magnetization is given by
\begin{align}
&M=\frac{1}{V}\int m(\bm{r})d\bm{r}, \\
&m(\bm{r})=\chi_s(\bm{q})B\cos(\bm{q}\cdot\bm{r}),
\end{align}
where the spin susceptibility $\chi_s(\bm{q})$ is obtained as
\begin{align}
\chi_s(\bm{q},\ef)=\frac{1}{V}\sum_{nm\bm{k}}\frac{-f_{n\bm{k}}+f_{m\bm{k}-\bm{q}}}{\e_{n\bm{k}}-\e_{m\bm{k}-\bm{q}}}\left|\langle n,\bm{k}|\frac{g\mb}{2}\s_z|m,\bm{k}-\bm{q}\rangle\right|^2,
\label{eq_kubo_cs}
\end{align}
where $V$ is the volume of the system, $f_{n\bm{k}}$ is the Fermi distribution function, $|n,\bm{k}\rangle$ is a Bloch state of the unperturbed Hamiltonian and $\e_{n\bm{k}}$ is its energy eigenvalue.

Taking the long wavelength limit $|\bm{q}|\to0$, we obtain 
\begin{align}
\lim_{|\bm{q}|\to0}\chi_s(\bm{q},\ef)=\ca(\ef)+\ce(\ef),
\end{align}
where $\ca(\ef)$ is the intraband contribution,
\begin{align}
\ca(\ef)=\frac{1}{V}\sum_{n\bm{k}}\left(-\frac{\partial f_{n\bm{k}}}{\partial\e_{n\bm{k}}}\right)\left|\langle n,\bm{k}|\frac{g\mb}{2}\s_z|n,\bm{k}\rangle\right|^2,
\label{eq_ca}
\end{align}
and $\ce(\ef)$ is the interband contribution,
\begin{align}
\ce(\ef)=\frac{1}{V}\sum_{n\neq m\bm{k}}\frac{-f_{n\bk}+f_{m\bk}}{\e_{n\bk}-\e_{m\bk}}\left|\langle n,\bm{k}|\frac{g\mb}{2}\s_z|m,\bm{k}\rangle\right|^2.
\label{eq_ce}
\end{align}
At the zero temperature, only electronic states on the Fermi surface contribute to $\ca$. On the other hand, all electronic states below the Fermi energy can contribute to $\ce$.
In order to get a finite $\ce$, the commutation relation between the Hamiltonian and the spin operator has to be non-zero,
\begin{align}
[H_0,\s_z]\neq0.
\end{align}
If the commutation relation is zero, the matrix elements in Eq.\ (\ref{eq_ce}) vanish and $\ce$ becomes zero. In the presence of the strong spin-orbit coupling, $\ce$ gives a large contribution.

%
%
%
%
%
%

\section{Massless Dirac fermions}
\label{sec_continuum}
We consider a model Hamiltonian for electrons in Dirac semimetals,
\begin{align}
H_{\rm continuum}=\hbar v\tau_z\bm{\s}\cdot\bk,
\label{eq_continuum_hami}
\end{align}
where $v$ is the velocity, $\bm{\s}$ and $\bm{\tau}$ are the triplets of Pauli matrices acting on the real spin and the pseudo spin (chirality) degrees of freedom. We calculate the spin susceptibility of the above model.
In the present model, the chirality is a good quantum number, so that the chirality degrees of freedom just double the spin susceptibility. The eigenstates of the Hamiltonian with positive chirality are given by
\begin{align}
|+,\bk\rangle&=\begin{pmatrix}
                         \cos\left(\t_{\bk}/2\right)e^{-i\phi_{\bk}/2} \\
                         \sin\left(\t_{\bk}/2\right)e^{i\phi_{\bk}/2}
                         \end{pmatrix}, \\
|-,\bk\rangle&=\begin{pmatrix}
                         -\sin\left(\t_{\bk}/2\right)e^{-i\phi_{\bk}/2} \\
                         \cos\left(\t_{\bk}/2\right)e^{i\phi_{\bk}/2}
                         \end{pmatrix},
\end{align}
where $|s,\bk\rangle$ is the eigenstate with the energy,
\begin{align}
\e_{s\bk}=s\hbar v k,
\end{align}
where $k=\sqrt{k_x^2+k_y^2+k_z^2}$ and $s=\pm1$. $\t_\bk$ and $\phi_\bk$ are the zenith and azimuth angles of the wave vector $\bk$.

The intraband and interband matrix elements are calculated as
\begin{align}
&\left|\langle s,\bk|\frac{g\mb}{2}\s_z|s,\bk\rangle\right|^2=\left(\frac{g\mb}{2}\right)^2\cos^2\t_\bk, \\
&\left|\langle -s,\bk|\frac{g\mb}{2}\s_z|s,\bk\rangle\right|^2=\left(\frac{g\mb}{2}\right)^2\sin^2\t_\bk.
\end{align}
We obtain an analytical expression for $\ca$,
\begin{align}
\ca(\ef)=\frac{1}{3\pi^2}\left(\frac{g\mb}{2}\right)^2\frac{\e_{\rm F}^2}{(\hbar v)^3},
\end{align}
where $\e_{\rm F}=\hbar v k_{\rm F}$ is the Fermi energy.
$\ca$ is proportional to the density of states $D(\e_{\rm F})$,
\begin{align}
D(\e_{\rm F})=\frac{\e_{\rm F}^2}{\pi^2(\hbar v)^3},
\end{align}
and corresponds to Pauli paramagnetism.
The interband contribution $\ce$ is also calculated analytically,
\begin{align}
\ce(\ef)=\frac{1}{3\pi^2}\left(\frac{g\mb}{2}\right)^2\frac{\ec^2-\e_{\rm F}^2}{(\hbar v)^3},
\end{align}
where $\ec=\hbar v \kc$ is a cutoff energy. This corresponds to the Van Vleck paramagnetism\cite{yu2010quantized,zhang2013topology}. In the present model, there are infinite states below the Fermi energy, so that we introduce a spherical cutoff with the radius $\kc$ in order to regularize the integration by $\bk$.

The spin susceptibility $\chi_s$, which is the sum of $\ca$ and $\ce$, is obtained as
\begin{align}
\chi_s(\ef)=\frac{1}{3\pi^2}\left(\frac{g\mb}{2}\right)^2\frac{\e_{\rm c}^2}{(\hbar v)^3}.
\label{eq_cs_continuum}
\end{align}
There are two important features. The spin susceptibility is independent of the Fermi energy \cite{koshino2016magnetic}, because the Fermi-energy dependent term of $\ca$ and $\ce$ exactly cancel each other. The spin susceptibility is proportional to $\ec^2$. In the present model, the cutoff energy $\ec$ is introduced by hand. Therefore the net value of the spin susceptibility can not be determined. At the first glance, this result is unreasonable, but we can appropriately determine the cutoff energy as we discuss in the next section.

\section{Lattice Model}
\label{sec_lattice}
In this section, we calculate the spin susceptibility of the Wilson-Dirac type cubic lattice model,
\begin{align}
H_{\rm Lattice}&=t\tau_z\sum_{i=x,y,z}\s_i \sin{k_ia}+m_\bk \tau_x, \notag \\
m_\bk&=m\sum_{i=x,y,z}(1-\cos k_ia),
\label{eq_lattice_hami}
\end{align}
where $\hbar v k_i{~}(i=x,y,z)$ in Eq.\ (\ref{eq_continuum_hami}) is simply replaced by $t\sin k_ia$ with the hopping energy $t$ and the lattice spacing $a$, and these parameters are related as
\begin{align}
\hbar v=ta.
\label{eq_ta}
\end{align}
The second term, $m_\bk\tau_x$, is introduced to gap out the point nodes other than the origin $(k_x,k_y,k_z)=(0,0,0)$.
In the vicinity of the origin, Eq.\ (\ref{eq_lattice_hami}) approximates the continuum model, Eq.\ (\ref{eq_continuum_hami}), within the first order of $k_i$. The eigenstates of the lattice model are given by
\begin{align}
|R,s,\bk\rangle&=\frac{1}{\sqrt{2\e_{s\bk}(\e_{s\bk}-t\sin k_za )}}
                        \begin{pmatrix}
                        t(\sin k_xa-i\sin k_ya) \\
                        \e_{s\bk}-t\sin k_za \\
                        0 \\
                        m_\bk
                        \end{pmatrix}, \\
|L,s,\bk\rangle&=\frac{1}{\sqrt{2\e_{s\bk}(\e_{s\bk}-t\sin k_za)}}
                        \begin{pmatrix}
                        -m_\bk \\
                        0 \\
                        -\e_{s\bk}+t\sin k_za \\
                        t(\sin k_xa+i\sin k_ya)
                        \end{pmatrix},
\end{align}
where $\e_{s\bk}=s\sqrt{t^2(\sin^2k_xa+\sin^2k_ya+\sin^2k_za)+m_\bk^2}$ and $s=\pm1$. $|R,s,\bk\rangle$ and $|L,s,\bk\rangle$ correspond to the eigenstates of the continuum model with positive and negative chiralities.

The intraband matrix elements are calculated as
\begin{align}
&\sum_{\alpha\beta s}\left(-\frac{\partial f_{s\bm{k}}}{\partial\e_{s\bm{k}}}\right)\left|\langle \alpha,s,\bk|\frac{g\mb}{2}\s_z|\beta,s,\bk\rangle\right|^2 \notag \\
&=\sum_s\left(-\frac{\partial f_{s\bm{k}}}{\partial\e_{s\bm{k}}}\right)\left(\frac{g\mb}{2}\right)^2\frac{2(t^2\sin^2k_za+m_\bk^2)}{\e_{s\bk}^2},
\label{eq_lattice_intra_matrix}
\end{align}
and the interband matrix elements are
\begin{align}
&\sum_{\alpha\beta s}\frac{-f_{s\bk}+f_{-s\bk}}{\e_{s\bk}-\e_{-s\bk}}\left|\langle \alpha,s,\bk|\frac{g\mb}{2}\s_z\right|\beta,-s,\bk\rangle|^2 \notag \\
&=(f_{-\bk}-f_{+\bk})\left(\frac{g\mb}{2}\right)^2\frac{2t^2(\sin^2k_xa+\sin^2k_ya)}{\e_{+\bk}^3}.
\label{eq_lattice_inter_matrix}
\end{align}
Using these matrix elements, we numerically calculate Eqs.\ (\ref{eq_ca}) and (\ref{eq_ce}).

\begin{figure}
\begin{center}
\leavevmode\includegraphics[width=0.9\hsize]{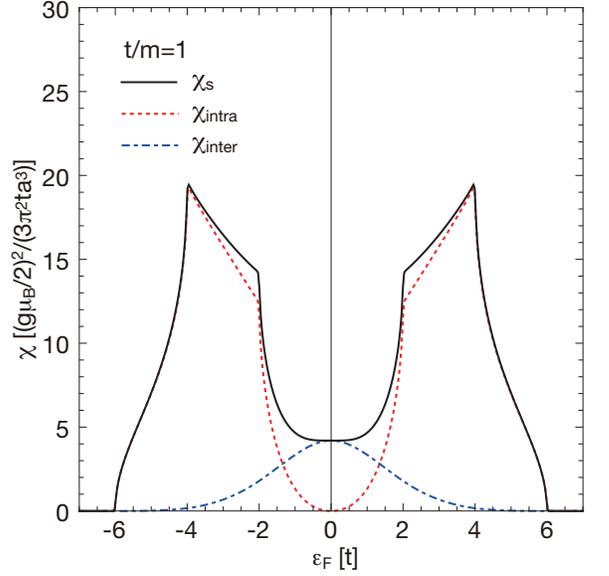}
\end{center}
\caption{The spin susceptibility of the lattice model as a function of the Fermi energy.
The solid black curve is the spin susceptibility $\chi_s$, the red dashed curve is the intraband contribution $\ca$, and the blue dashed curve is the interband contribution $\ce$.
}
\label{fig_sus_ef}
\end{figure}

Figure \ref{fig_sus_ef} shows the spin susceptibility as a function of the Fermi energy $\ef$. Around the zero energy where the dispersion relation is linear, the qualitative behavior of the spin susceptibility of the lattice model is the same as the continuum model. The interband contribution has a peak structure at the zero energy. The width of the peak is related to the structure of the Hamiltonian. The Hamiltonian is composed of two terms, the first sin term, which does not commute with the spin operator,
\begin{align}
&H_{\rm s}=t\tau_z\sum_{i=x,y,z}\s_i\sin k_ia, \\
&[H_{\rm s},\s_z]\neq0,
\end{align}
and the second cos term, which commutes with the spin operator,
\begin{align}
&H_{\rm c}=m\tau_x\sum_{i=x,y,z}(1-\cos k_ia), \\
&[H_{\rm c},\s_z]=0.
\end{align}
Around the Dirac point, the electronic states are approximately described by $H_{\rm s}$, and the interband matrix element is finite. Far from the Dirac point, on the other hand, the electronic states are approximately described by $H_{\rm c}$, and the interband matrix element is negligibly small. Therefore, the interband contribution has the peak structure and finite value near the zero energy.
The peak decays when the cos term $H_{\rm c}$ is comparable to the sin term $H_{\rm s}$.

\begin{figure}
\begin{center}
\leavevmode\includegraphics[width=0.9\hsize]{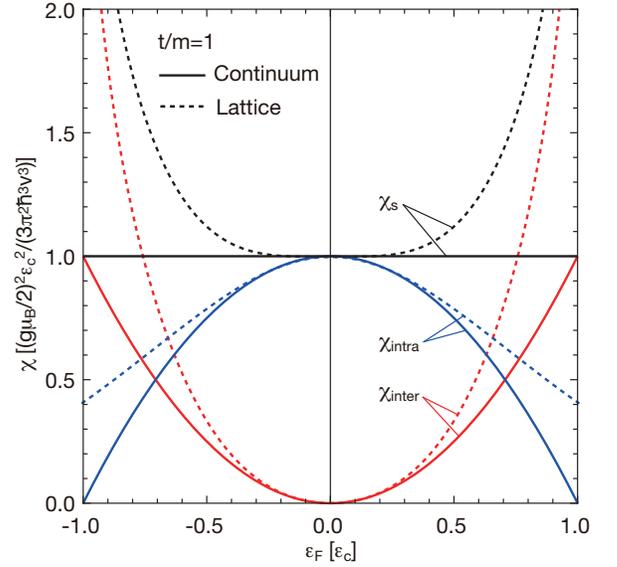}
\end{center}
\caption{The spin susceptibility of the continuum model (solid curves) and the lattice model (dashed curves) as a function of the Fermi energy.
}
\label{fig_sus_ef_comp}
\end{figure}

Here, we relate the peak width of $\chi_{\rm inter}$ and the cutoff energy $\e_{\rm c}$, which is introduced in the previous section. In the continuum model, the interband contribution vanishes at the cutoff energy, while in the lattice model, the interband contribution decays far from the Dirac point.
Therefore, we assume that the cutoff energy corresponds to the peak width and is determined by
\begin{align}
t\sin (\kc a/f)=m\left[1-\cos (\kc a/f)\right],
\end{align}
which means the sin term and the cos term is comparable.
In the above equation, we introduce a numerical factor $f$ to fit the spin susceptibility of the continuum and lattice model as discussed following.
Solving the above equation, we obtain
\begin{align}
\kc a&=2f\arctan\left(\frac{t}{m}\right).
\label{eq_kca}
\end{align}
In Fig.\ \ref{fig_sus_ef_comp}, we compare the spin susceptibility of the continuum model and the lattice model. Using Eqs.\ (\ref{eq_ta}) and (\ref{eq_kca}), the two spin susceptibilities are compared in the same unit. The numerical factor $f$ is determined as
\begin{align}
f\simeq1.305,
\end{align}
to get quantitative agreement between the two spin susceptibilities at $\ef=0$.
In the vicinity of the zero energy, they are good agreement with each other. On the other hand, we see the deviation apart from the zero energy because of the deviation from the liner dispersion relation.

\begin{figure}
\begin{center}
\leavevmode\includegraphics[width=0.8\hsize]{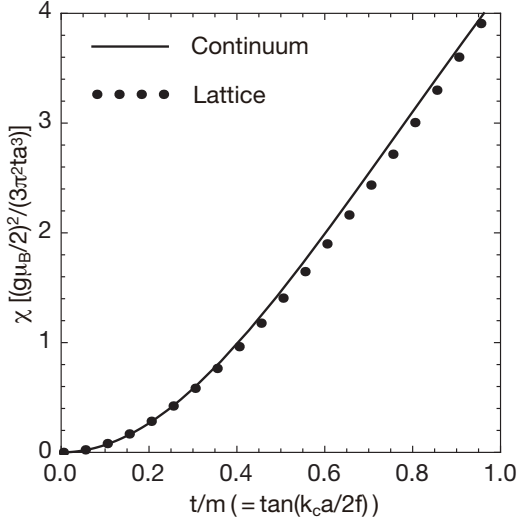}
\end{center}
\caption{The spin susceptibility at $\e_{\rm F}=0$ as a function of $t/m$. The solid curve represents the continuum model and the dotted curve represents the lattice model.
}
\label{fig_sus_kc}
\end{figure}

Figure\ \ref{fig_sus_kc} compares the spin susceptibility of the continuum model and that of the lattice model at $\ef=0$ as a function of $t/m$. Again we see the quantitative agreement between the two spin susceptibilities.
In a condition that $t/m\ll1$, we can derive an approximate analytical expression for the spin susceptibility of the lattice model.
In this condition, the interband matrix elements Eq.\ (\ref{eq_lattice_inter_matrix}) is expanded as
\begin{align}
\frac{2t^2(\sin^2k_xa+\sin^2k_ya)}{\e_{+\bk}^3}\simeq\frac{2(ta)^2(k_x^2+k_y^2)}{\left[(tak)^2+(ma^2k^2/2)^2\right]^{3/2}},
\end{align}
and the spin susceptibility of the lattice model is calculated as
\begin{align}
\chi_s(\ef=0)&\simeq\frac{1}{(2\pi)^3}\int_0^\infty k^2dk\int_0^\pi\sin\t d\t\int_0^{2\pi}d\phi \notag \\
                      &\hspace{1cm}\times\frac{2(ta)^2(k_x^2+k_y^2)}{\left[(tak)^2+(ma^2k^2/2)^2\right]^{3/2}} \notag \\
                    &=\frac{8}{3\pi^2}\frac{t^2}{m^2a^2}.
\end{align}
In the present approximation, Eq.\ (\ref{eq_kca}) becomes $\kc a\simeq2f(t/m)$. Consequently, we obtain $\chi_s(\ef=0)\propto \kc^2$. This is consistent with the above agreement.

\section{Massive Dirac Fermions}
\label{sec_ti}
%
%
%
%
%
%
%
In this section, we calculate the spin susceptibility of the massive Dirac Hamiltonian, which can describes an electronic state of topological insulators. A magnetically doped topological insulator is one of the candidate materials for magnetic Weyl semimetals \cite{yu2010quantized,cho2011possible,liu2013chiral,kurebayashi2014weyl}. Therefore, to clarify the properties of the spin susceptibility of topological insulators is an important issue to realize magnetic Weyl semimetals.

The electronic state is described by the effective Hamiltonian \cite{zhang2009topological,liu2010model},
\begin{align}
H_0=\e_{\bm k}+M_{\bm k}\tau_z+B_0\tau_y k_z+A_0(\tau_x\s_x k_y-\tau_x\s_y k_x),
\end{align}
where $\e_{\bm k}=C_0+C_1k_z^2+C_2k_{\|}^2$, $M_{\bm k}=M_0+M_1k_z^2+M_2k_{\|}^2$, and $k_\|=\sqrt{k_x^2+k_y^2}$. In the following calculation, the parameters are taken as $C_0=-0.0083[{\rm eV}], C_1=5.74[{\rm eV\AA^2}], C_2=30.4[{\rm eV\AA^2}], M_1=6.86[{\rm eV\AA^2}], M_2=44.5[{\rm eV\AA^2}], A_0=3.33[{\rm eV\AA}]$, and $B_0=2.26[{\rm eV\AA}]$, which are the parameters for the topological insulator ${\rm Bi}_2{\rm Se}_3$ \cite{liu2010model}. The above Hamiltonian describes ordinary insulators, Dirac semimetals, and topological insulators by tuning the parameter $M_0$, which is related to the strength of the spin-orbit coupling. In the presence of a magnetic field, the Zeeman coupling is given by
\begin{align}
H_{\rm Zeeman}
                           =-\bm{M}^{\rm spin}\cdot\bm{B},
\label{eq_ti_zeeman}
\end{align}
where the spin operators, $\bm{M}^{\rm spin}$, are written as
\begin{align}
    M^{\rm spin}_x=&\frac{\mb}{2}(g_{xy+}\s_x+g_{xy-}\tau_z\s_x), \\
    M^{\rm spin}_y=&\frac{\mb}{2}(g_{xy+}\s_y+g_{xy-}\tau_z\s_y), \\
    M^{\rm spin}_z=&\frac{\mb}{2}(g_{z+}\s_z+g_{z-}\tau_z\s_z).
\label{eq_ti_spin}
\end{align}
We set the effective $g$ factors as $g_{z+}=10.65$, $g_{z-}=14.75$, $g_{xy+}=-0.34$, and $g_{xy-}=4.46$, which are also the parameters for the topological insulator ${\rm Bi}_2{\rm Se}_3$ \cite{liu2010model}.
In this model, there are two kinds of Zeeman terms, "orbital-independent" term ($\s_\alpha$) and "orbital-dependent" term ($\tau_z\s_\alpha$) \cite{nakai2016crossed}. This originates from the non-equality of the effective $g$ factors in the two orbitals.
The eigenstates of the above Hamiltonian are given by
\begin{align}
|1,+,\bk\rangle&=\frac{1}{\sqrt{2\e_{+\bk}(\e_{+\bk}+M_\bk)}}
                        \begin{pmatrix}
                        \e_{+\bk}+M_\bk \\
                        0 \\
                        iB_0k_z \\
                        -iA_0k_+
                        \end{pmatrix}, \\
|2,+,\bk\rangle&=\frac{1}{\sqrt{2\e_{+\bk}(\e_{+\bk}+M_\bk)}}
                        \begin{pmatrix}
                        0 \\
                        \e_{+\bk}+M_\bk \\
                        iA_0k_- \\
                        iB_0k_z
                        \end{pmatrix}, \\
|1,-,\bk\rangle&=\frac{1}{\sqrt{2\e_{-\bk}(\e_{-\bk}-M_\bk)}}
                        \begin{pmatrix}
                        -iB_0k_z \\
                        -iA_0k_+ \\
                        \e_{-\bk}-M_\bk \\
                        0
                        \end{pmatrix}, \\                        
|2,-,\bk\rangle&=\frac{1}{\sqrt{2\e_{-\bk}(\e_{-\bk}-M_\bk)}}
                        \begin{pmatrix}
                        iA_0k_- \\
                        -iB_0k_z \\
                        0 \\
                        \e_{-\bk}-M_\bk
                        \end{pmatrix},
\end{align}
where $k_\pm=k_x\pm i k_y$, 
and the energy for $|n,s,\bk\rangle$ is given by
\begin{align}
\e_{s\bk}=s\sqrt{A_0^2k^2_{\|}+B_0^2k_z^2+M_\bk^2}.
\end{align}

Based on the symmetry, the spin susceptibility along the $x$ axis and the $y$ axis exhibit the same behavior. Therefore, we calculate the spin susceptibility along the $x$ axis and the $z$ axis.
The intraband matrix elements are calculated as
\begin{align}
&\sum_{nm}\left(-\frac{\partial f_{s\bm{k}}}{\partial\e_{s\bm{k}}}\right)|\langle n,s,\bk|M_x^{\rm spin}|m,s,\bk\rangle|^2=\left(-\frac{\partial f_{s\bm{k}}}{\partial\e_{s\bm{k}}}\right)\left(\frac{\mb}{2}\right)^2 \notag \\
&\hspace{-0.5cm}\times\frac{2\left[g_{xy+}^2(\e_{s\bk}^2-A_0^2k_x^2)+2g_{xy+}g_{xy-}\e_{s\bk}M_\bk+g_{xy-}^2(A_0^2k_x^2+M_\bk^2)\right]}{\e_{s\bk}^2},
\label{eq_intra_x}
\end{align}
and
\begin{align}
&\sum_{nm}\left(-\frac{\partial f_{s\bm{k}}}{\partial\e_{s\bm{k}}}\right)|\langle n,s,\bk|M_z^{\rm spin}|m,s,\bk\rangle|^2=\left(-\frac{\partial f_{s\bm{k}}}{\partial\e_{s\bm{k}}}\right)\left(\frac{\mb}{2}\right)^2 \notag \\
&\hspace{-0.2cm}\times\frac{2\left[g_{z+}^2(B_0^2k_z^2+M_\bk^2)+2g_{z+}g_{z-}\e_{s\bk}M_\bk+g_{z-}^2(\e^2_{s\bk}-B_0^2k_z^2)\right]}{\e_{s\bk}^2}.
\label{eq_intra_z}
\end{align}
The interband matrix elements are
\begin{align}
&\sum_{nms}\frac{-f_{s\bk}+f_{-s\bk}}{\e_{s\bk}-\e_{-s\bk}}|\langle n,s,\bk|M_x^{\rm spin}|m,-s,\bk\rangle|^2 \notag \\
&=(f_{-\bk}-f_{+\bk})\left(\frac{\mb}{2}\right)^2\frac{2\left[g_{xy+}^2A_0^2k_x^2+g_{xy-}^2(A_0^2k_y^2+B_0^2k_z^2)\right]}{\e_{+\bk}^3},
\end{align}
and
\begin{align}
&\sum_{nms}\frac{-f_{s\bk}+f_{-s\bk}}{\e_{s\bk}-\e_{-s\bk}}|\langle n,s,\bk|M_z^{\rm spin}|m,s,\bk\rangle|^2 \notag \\
&=(f_{-\bk}-f_{+\bk})\left(\frac{\mb}{2}\right)^2\frac{2\left[g_{z+}^2A_0^2(k_x^2+k_y^2)+g_{z-}^2B_0^2k_z^2\right]}{\e_{+\bk}^3}.
\end{align}

\begin{figure*}
\begin{center}
\leavevmode\includegraphics[width=1\hsize]{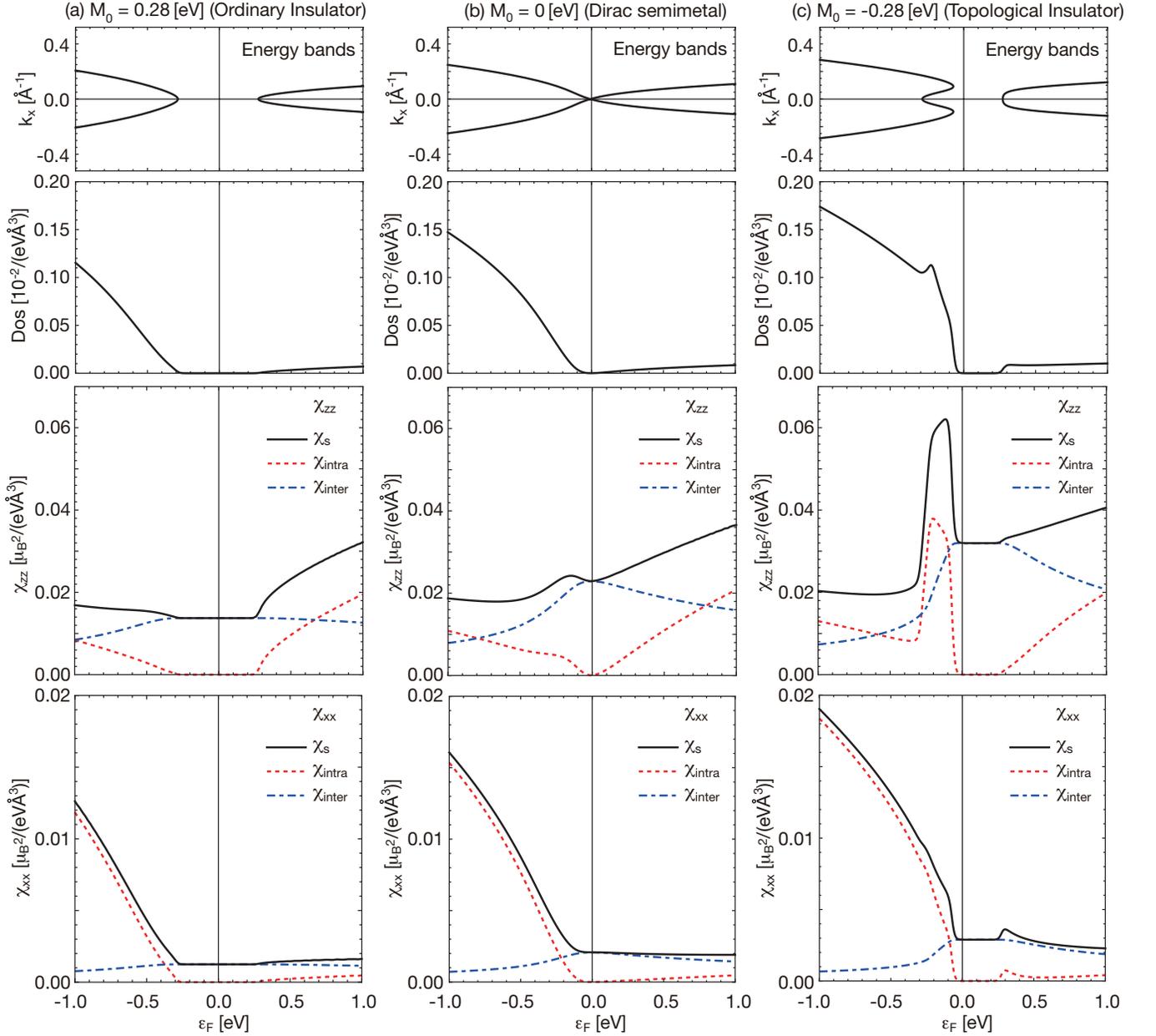}
\end{center}
\caption{The density of states and the spin susceptibility of (a) Ordinary insulator, (b) Dirac semimetal, and (c) Topological insulator as a function of the Fermi energy. The top panels show the energy bands, where we set $k_y=k_z=0$.
}
\label{fig_sus_ti}
\end{figure*}

\begin{figure}
\begin{center}
\leavevmode\includegraphics[width=0.8\hsize]{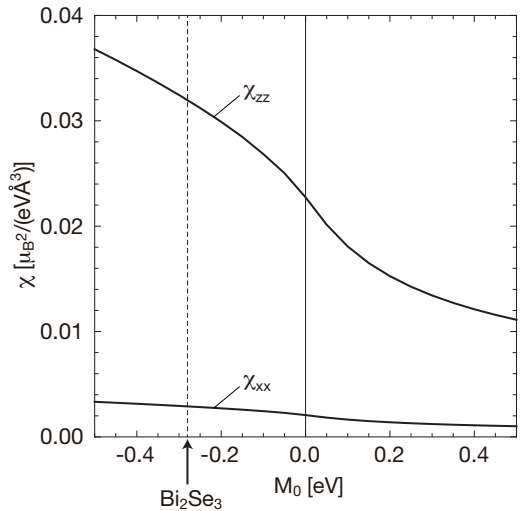}
\end{center}
\caption{The spin susceptibility of the $x$ and $z$ directions in the energy gap as a function of $M_0$.
The vertical dashed line corresponds to the value of $M_0$ for ${\rm Bi}_2{\rm Se}_3$.
}
\label{fig_sus_ti_m0}
\end{figure}

The spin susceptibility is numerically calculated in a similar manner to the previous sections. Figure\ \ref{fig_sus_ti} shows the density of states and the spin susceptibility as a function of the Fermi energy $\ef$. The top panels in Fig.\ \ref{fig_sus_ti} show the energy bands. We calculate them for three parameters (a) $M_0=0.28[{\rm eV}]$ (Ordinary insulator), (b) $M_0=0.0[{\rm eV}]$ (Dirac semimetal), and (c) $M_0=-0.28[{\rm eV}]$ (Topological insulator). Even in the current effective model, which includes the anisotropy and the two types of the Zeeman term, the qualitative behavior of the interband contribution is similar to that of the previous models. The interband contribution takes the maximum value in the energy gap or at the band touching point, where the density of states vanishes. Away from the zero energy, the interband contribution monotonically decreases
in a similar manner to the previous model. On the other hand,
the intraband contribution behaves in a slightly different manner from the precious model.
In the previous models, the intraband contribution is proportional to the density of states. In the current model, the density of states of the valence band is larger than the conduction band, but the intraband contributions for $\chi_{zz}$ in the valence and conduction bands are comparable. This originates from the cross term of $g_{z+}$ and $g_{z-}$ in Eq.\ (\ref{eq_intra_z}). The cross term gives positive contribution in the conduction band and negative in the valence band. Consequently, the intraband contribution in the valence and conduction bands are comparable. On the other hand, the intraband contributions for $\chi_{xx}$ in the valence and conduction bands are not comparable. This is because the effective $g$ factors $g_{xy+}$ and $g_{xy-}$ have opposite signs, so that the cross term does not work as the case of $\chi_{zz}$, where $g_{z+}$ and $g_{z-}$ have same signs.
In Fig.\ \ref{fig_sus_ti} (c) the topological insulator case, there is another important feature. The intraband contribution for $\chi_{zz}$ exhibits a peak structure in the valence band. The peak width corresponds to the band inverted region. On the other hand, there is no peak structure in $\chi_{xx}$.

In Fig.\ \ref{fig_sus_ti_m0}, we plot the spin susceptibility in the energy gap as a function of $M_0$. The spin susceptibility increases with the decrease of $M_0$, which means the increase of the spin-orbit coupling \cite{yu2010quantized,zhang2013topology}. The strong spin-orbit coupling gives the large interband contribution. $\chi_{zz}$ is much larger than $\chi_{xx}$, because the effective $g$ factors for the $z$ direction are much larger than the $x$ direction.

\section{Conclusion}
\label{sec_conclusion}

We have studied the spin susceptibility of the Dirac semimetals. The spin susceptibility is calculated for the massless Dirac continuum model and the Wilson-Dirac lattice model. In the massless Dirac continuum model, we have to introduce the cutoff energy $\e_{\rm c}$ in order to regularize the integration. The spin susceptibility is independent of the Fermi energy $\ef$ and proportional to $\e_{\rm c}^2$. We find that the cutoff energy is appropriately determined and related to the some parameters of the lattice model. The cutoff energy corresponds to the energy where the band dispersion deviates from the linear dispersion relation. The spin susceptibility of the lattice model is in quantitatively good agreement with the massless Dirac continuum model. We also calculate the spin susceptibility of massive Dirac fermions with the Zeeman coupling including the orbital dependent term and orbital independent term. The spin susceptibility along the $z$ axis is enhanced in the conduction band because of the existence of two types of the Zeeman term and has the peak structure in the band inverted region, which are not observed in the spin susceptibility along the $x$ axis.

\section*{ACKNOWLEDGMENT}
The authors thank Yasufumi Araki and Masaki Oshikawa for helpful discussions.
This work was supported by Kakenhi Grants-in-Aid (Nos. JP15H05854 and JP17K05485) from the Japan Society for the Promotion of Science (JSPS).

\bibliography{spin_susceptibility}

\end{document}